\documentstyle[12pt]{article}

\newcommand {\nn}    {\nonumber}
\newcommand {\vs}[1]  { \vspace*{#1 cm} }

\newcounter{eq}
\newcounter{sc}

\newcommand {\MPL}  {Mod.Phys.Lett.}
\newcommand {\NP}   {Nucl.Phys.}
\newcommand {\PL}   {Phys.Lett.}
\newcommand {\PR}   {Phys.Rev.}
\newcommand {\PRL}   {Phys.Rev.Lett.}
\newcommand {\CMP}  {Comm.Math.Phys.}
\newcommand {\AP}   {Ann.of Phys.}

\newcommand {\CQG}  {Class.Quant.Grav.}
\newcommand {\LMP}  {Lett.Math.Phys.}

\def\overleftrightarrow#1{\vbox{\ialign{##\crcr
 $\leftrightarrow$\crcr\noalign{\kern-1pt\nointerlineskip}
 $\hfil\displaystyle{#1}\hfil$\crcr}}}

\newlength{\minitwocolumn}
\begin{document}


\begin{flushright}
EDO-EP-17\\
January, 1998\\
\end{flushright}
\vspace{30pt}

\pagestyle{empty}
\baselineskip15pt

\begin{center}
{\large\bf Background Independent  Matrix Models 
\vskip 1mm}

\vspace{20mm}

Ichiro Oda
          \footnote{
          E-mail address:\ ioda@edogawa-u.ac.jp
                  }
\\
\vspace{10mm}
          Edogawa University,
          474 Komaki, Nagareyama City, Chiba 270-01, JAPAN \\

\end{center}


\vspace{15mm}
\begin{abstract}
A class of background independent matrix models is made for
which the structure of both local gauge symmetries and classical
solutions is clarified. These matrix models do not involve
a space-time metric and provide the matrix analogs of topological
Chern-Simons and BF theories. It is explicitly shown that 
the BF type of matrix model can be formulated in any space-time 
dimension and include 3+1 dimensional gravity as a 
special case.  Moreover, 
we discuss some generalization of the model to include a fermionic 
BRST-like symmetry whose partition function is related to the 
Casson invariant. 

\vspace{15mm}

\end{abstract}

\newpage
\pagestyle{plain}
\pagenumbering{arabic}


\rm
\section{Introduction}

One of the most important observations in the recent developments
of string theory is that D-instanton and D-particle \cite{Pol}
may be the microscopic degrees of freedom of IIB superstring 
theory \cite{IKKT} and M-theory \cite{M}, respectively. In both 
the theories, the space-time coordinates are expressed in terms of 
$N \times N$ hermitian matrices describing coordinates of
D-instanton or D-particle. For instance, M-theory in the infinite 
momentum frame is expected to be equivalent to a quantum mechanics of
$U(N)$ matrices in the $N \rightarrow \infty$, with the Hamiltonian
that comes from the one dimensional reduction of ten dimensional
super Yang-Mills theory \cite{M}. This feature of the space-time
coordinates as matrices yields a totally new interpretation
about the space-time structure.
Namely, the conventional description of the space-time as a 
continuous manifold is in itself meaningful only in the long 
distance region where the space-time coordinates are commutable 
and diagonal while the space-time is quantized and has a 
discretized structure in the short distance regime.

Despite such impressive developments of matrix models, it is 
fair to say that we are still far from having a complete understanding 
of non-perturbative formulation of string theory. In particular,
a big mystery is the problem of background independence. 
In the matrix models constructed so far \cite{IKKT, M}, it is 
always assumed space-time background to be flat. Since the matrix
theories must involve a theory of quantum gravity, they are not
allowed to have their most fundamental formulation in a fixed
classical space-time manifold and the space-time geometry should
emerge from a more fundamental theory that is independent of 
background metric. 

Closely related to the background independence is that we have no
clear understanding of how the matrix theories are connected with
Einstein's general relativity. Even if there is circumstancial
evidence that the low energy theory of the matrix models contains
general relativity, it is quite obscure how general relativity is
derived from the matrix models in a comprehensive manner 
\cite{Susskind}. 
These two big questions also have a deep connection with 
the crucial question of what the underlying gauge symmetry and 
the fundamental principle behind the matrix theories are. 
   
In the present paper, we would like to address the question of
whether a background independent formulation of the matrix
model is possible. The main idea of this article is to
construct the matrix models of topological Chern-Simons 
\cite{Witten1} and BF theories \cite{BF1, BF2, BF3, BF4}. 
As a consequence, the matrix models obtained 
in this way do not depend on background metric and contain
the local translation invariance in a manifest way. 
Incidentally, a different type of the matrix models has
been already made on a basis of the topological quantum 
field theory \cite{Hirano, Oda}. 

The paper is organized as follows. In section 2 we study
Chern-Simons type of background independent matrix model 
that was originally introduced by Smolin \cite{Smolin} 
and examine some intriguing problems such as 
its classical solutions and canonical formalism. 
In section 3, we construct a new background 
independent matrix model based on topological BF theory.
In contrast with the Chern-Simons type, this new matrix
model is not only formulated in any space-time dimension
but also yields general relativity reduced to a point
in 2, 3 and 4 dimensions by selecting appropriate
classical solutions. In section 4, we incorporate the spinors in
the above theory and construct a new matrix model with 
BRST-like supersymmetry whose partition function yields
the Casson invariant \cite{Witten2}. 
The final section is devoted to discussions.

\section{ The Chern-Simons matrix model}

In this section, we shall not only review the Chern-Simons type of
matrix model which was originally introduced in \cite{Smolin},
but also examine its classical solutions and canonical formalism.

Let us consider a simple game constructing the action which 
consists of only the hermitian matrices 
$X_\mu (\mu=0,1, \cdots, D-1)$ and is independent 
of the background metric. Almost a unique answer 
is to just line up all the $X_\mu$'s, take the trace 
of them and then contract 
the $D$ indices by the Levi-Civita tensor density 
$\varepsilon^{\mu_1 \mu_2 \cdots \mu_D}$ to make 
a c-number scalar. As a consequence, we obtain the topological
matrix model
\begin{eqnarray}
S_{CS}^D = \varepsilon^{\mu_1 \mu_2 \cdots \mu_D} 
Tr X_{\mu_1} X_{\mu_2} \cdots X_{\mu_D}.
\label{2.1}
\end{eqnarray}
Interestingly enough, we can construct such an action only in 
the case that $D$ is odd numbers since the action with even numbers
of $X_\mu$ is identically zero by the following identity:
\begin{eqnarray}
S_{CS}^D &=& \varepsilon^{\mu_1 \mu_2 \cdots \mu_D} 
Tr X_{\mu_1} X_{\mu_2} \cdots X_{\mu_D} \nn\\
&=& \varepsilon^{\mu_1 \mu_2 \cdots \mu_D} 
Tr X_{\mu_D} X_{\mu_1} \cdots X_{\mu_{D-1}} \nn\\
&=& (-1)^{D-1} \varepsilon^{\mu_D \mu_1 \mu_2 \cdots \mu_{D-1}} 
Tr X_{\mu_D} X_{\mu_1} \cdots X_{\mu_{D-1}} \nn\\
&=& (-1)^{D-1} S_{CS}^D,
\label{2.2}
\end{eqnarray}
where we have used the cyclic property of trace and the totally
antisymmetric property of the Levi-Civita tensor density. Thus
we will set $D$ to be $2d+1$ with $d \in {\bf Z_+} \cup \{0\}$
in this section.
Incidentally, the topological matrix model with any number
of $X_\mu$ will be built in the next section.

The equations of motion derived from the action (\ref{2.1}) read
\begin{eqnarray}
\varepsilon^{\mu \mu_1 \mu_2 \cdots \mu_{2d}} 
X_{\mu_1} X_{\mu_2} \cdots X_{\mu_{2d}} = 0.
\label{2.3}
\end{eqnarray}
Note that (\ref{2.3}) does not include the metric tensor
in comparison with the equations of motion derived from
IIB and M matrix models \cite{IKKT, M} 
whose formal expression is provided by
\begin{eqnarray}
\eta^{\mu\nu} \left[ X_\mu, \left[ X_\nu, X_\rho \right] \right]
= 0
\label{2.4}
\end{eqnarray}
with the flat Minkowskian metric $\eta^{\mu\nu} = diag(-1, +1, \cdots,
+1)$. At this stage, it is useful to find the classical solutions 
satisfying
the equations of motion (\ref{2.3}). One obvious solution is the
one satisfying the equation $\left[ X_\mu, X_\nu \right]
= 0$, that is, this solution has the form of the diagonal $N \times N$
matrix
\begin{eqnarray}
 X_\mu =  \pmatrix{
X_\mu^{(1)} & {} & {} & {} \cr
{}          & {} & \ddots & {} \cr
{}          & {} &  {}    & X_\mu^{(N)} \cr
},
\label{2.5}
\end{eqnarray}
which we call "classical space-time" in this paper.  Next nontrivial
solution is "string" solution given by
\begin{eqnarray}
 X_\mu = \left( X_0, X_1, 0, \cdots, 0 \right),
\label{2.6}
\end{eqnarray}
where we have considered the string along 1st axis without
losing generality. Similarly, "membrane" solution stretched out
in the direction of 1st and 2nd axes reads
\begin{eqnarray}
 X_\mu = \left( X_0, X_1, X_2, 0, \cdots, 0 \right).
\label{2.7}
\end{eqnarray}
It is obvious that this kind of solutions continues to exist until
"$(2d-1)$-brane"
\begin{eqnarray}
 X_\mu = \left( X_0, X_1, X_2, \cdots, X_{2d-1}, 0 \right).
\label{2.8}
\end{eqnarray}
Moreover, a solution associated with several "$k$-branes"
$(1 \le k \le 2d-1)$ can be built out of the above solution
for single "$k$-brane" in a perfectly similar way to
the case of IIB matrix model \cite{IKKT}.
For instance, the solution for two "strings" separated 
by the distance $b$ along 2nd axis is given by
\begin{eqnarray}
X_0 = \pmatrix{
x_0 & 0   \cr
0   & x_0 \cr }, \ 
X_1 = \pmatrix{
x_1 & 0   \cr
0   & x_1 \cr }, \nn\\
X_2 = \pmatrix{
\frac{b}{2} & 0   \cr
0   & -\frac{b}{2} \cr }, \ 
X_3 = \cdots = X_{2d} = 0, 
\label{2.9}
\end{eqnarray}
where $x_0$ and $x_1$ are certain nonzero elements.
Of course, the specific choice of a form of the matrix $X_\mu$
leads to other classical solutions of the equations of 
motion (\ref{2.3}), but I could not find any physical importance
on them as solutions of Matrix Theory.

Now let us turn our attention to the symmetries in the action
(\ref{2.1}). It is remarkable that as well as the conventional
gauge symmetry
\begin{eqnarray}
 X_\mu \rightarrow X_\mu^\prime = U X_\mu U^{-1}
\label{2.10}
\end{eqnarray}
with $U \in U(N)$, the action (\ref{2.1}) is invariant under
the local translation of the diagonal element
\begin{eqnarray}
 X_\mu \rightarrow X_\mu^\prime =  X_\mu + V_\mu(X) \ {\bf 1}
\label{2.11}
\end{eqnarray}
with $V_\mu(X)$ being not a matrix but a c-number function of $X_\mu$.
This symmetry is in sharp contrast with the matrix models 
\cite{IKKT, M} where $V_\mu$ is a global parameter of c-number.
In other words, the global translation in \cite{IKKT, M} is now
promoted to the local translation. In this respect,  
it is of interest to recall the following things. 
Firstly, in the matrix models
\cite{IKKT, M} the diagonal matrix like (\ref{2.5}) corresponds to  
the classical space-time coordinates while the non-diagonal
matrix describes the interactions. Hence the local symmetry
(\ref{2.11}) coincides with the local space-time translation 
at the classical level. Secondly, it is well known that general
relativity is the gauge theory with the local translation as
the gauge symmetry, so the existence of this symmetry might be
a signal of the existence of general relativity in this matrix
model though we need more studies to confirm this conjecture
in future. 

In the remainder of this section, we would like to consider the 
possibility of deriving the gauge symmetries (\ref{2.10}) 
and (\ref{2.11}) in the canonical formalism. In order to
tame the action (\ref{2.1}) in the canonical framework,
it is necessary to introduce a fictitious time $\tau$ 
into the theory and
assume that $X_\mu$'s are function of $\tau$. The idea
is then to consider a similar (but different) action to
(\ref{2.1}), from which to gain useful information 
about constraints of the action (\ref{2.1}) through
the canonical formalism of the similar action. 
As such a deformed action, let us consider 
\begin{eqnarray}
I_{CS}^{2d+1} = \int d\tau \ \varepsilon^{\mu_1 \mu_2 \cdots 
\mu_{2d+1}} Tr \left({\it D_\tau} X_{\mu_1} \right)
X_{\mu_2} \cdots X_{\mu_{2d+1}},
\label{2.12}
\end{eqnarray}
where the covariant derivative ${\it D}_\tau X_\mu
= \partial_\tau X_\mu + \left[ A_\tau, X_\mu \right]$
with a fictitious gauge field $A_\tau$ is introduced.
Note that this action (\ref{2.12}) reduces simply 
to the original action (\ref{2.1}) at the boundary
in a gauge with $A_\tau = 0$. This is the reason why
we have chosen the action (\ref{2.12}) in order to
implement the canonical analysis of the action 
(\ref{2.1}). 

The canonical conjugate momenta corresponding to
$X_\mu$ are given by
\begin{eqnarray}
P^\mu = \varepsilon^{\mu \mu_1 \mu_2 \cdots 
\mu_{2d}} X_{\mu_1} X_{\mu_2} \cdots X_{\mu_{2d}},
\label{2.13}
\end{eqnarray}
from which $Tr P^\mu = 0$ holds identically. On the
other hand, the canonical conjugate momentum $\pi$
corresponding to $A_\tau$ vanishes trivially 
since the action (\ref{2.12})
does not involve the kinetic term for $A_\tau$.
The Hamiltonian {\it H} can be easily calculated to
\begin{eqnarray}
{\it H} = - \varepsilon^{\mu_1 \mu_2 \cdots 
\mu_{2d+1}} Tr A_\tau \left[ X_{\mu_1},
X_{\mu_2} \cdots X_{\mu_{2d+1}} \right].
\label{2.14}
\end{eqnarray}
Thus, the condition that the time evolution of the primary
constraint $\pi \approx 0$ also vanishes weakly under this
Hamiltonian gives rise to the secondary constraint 
\begin{eqnarray}
\left[ X_\mu, P^\mu \right] \approx 0,
\label{2.15}
\end{eqnarray}
which is nothing but Gauss's law constraint. In terms of this
Gauss's law constraint, the Hamiltonian (\ref{2.14}) becomes
weakly zero so that no more constraint occurs.
Therefore all the constraints are summarized to be in the form
\begin{eqnarray}
\phi^\mu &\equiv& P^\mu - \varepsilon^{\mu \mu_1 \mu_2 \cdots 
\mu_{2d}} X_{\mu_1} X_{\mu_2} \cdots X_{\mu_{2d}}
\approx 0, \nn\\
\chi &\equiv& \left[ X_\mu, P^\mu \right] \approx 0,
\label{2.16}
\end{eqnarray}
where the constraint $\pi \approx 0$ is excluded from this
constraint system by picking
a gauge with $A_\tau = 0$. Then it is straightforward to
check that these constraints (\ref{2.16}) are the 
first-class contraints so that they generate the infinitesimal
gauge transformations. Actually, we can easily see that
$\phi^\mu \approx 0$ generates the topological symmetry
$X_\mu \rightarrow X_\mu + \varepsilon_\mu(X)$ and
$\chi \approx 0$ does the usual gauge transformation.
The reason why we have the topological symmetry is quite
simple. This is because as suggested above 
in the gauge with $A_\tau=0$ the action (\ref{2.12}) 
precisely reduces to a surface term. The role of
the fictitious gauge field $A_\tau$ is just to isolate
the usual gauge symmetry from the topological symmetry.

So far we have developed the canonical formalism for the 
deformed action (\ref{2.12}). We are now in a position to
extract the information about the first-class constraints 
describing the gauge symmetries
of the action (\ref{2.1}) from the constraints (\ref{2.16})
of the action (\ref{2.12}).
All we have to do is to take the contraints with the forms 
of $Tr \phi^\mu = Tr p^\mu \approx 0$ plus $\chi \approx 0$
from (\ref{2.16}). 
It is then obvious that the algebra closes among these constraints
and these constraints generate
the gauge symmetries (\ref{2.10}) and (\ref{2.11}).

\section{ The BF matrix model }

In the previous section, we have considered the Chern-Simons
matrix model, but this model has some problems. In particular,
it is quite unsatisfactory that we cannot construct the 
matrix model in even space-time dimensions. Furthermore,
it seems to be difficult to make a supersymmetric extension of the
Chern-Simons matrix model without introducing the background
metric. Finally, it is at present unclear that the Chern-Simons matrix 
model has a relationship with general gravity. Luckily, we
have already met a similar situation to this in topological quantum
field theories where the Chern-Simons theory \cite{Witten1}
is replaced with the BF theory \cite{BF1, BF2, BF3, BF4} 
in order to overcome these impasse.
In the case of the matrix model at hand we also proceed with 
the same line of argument as the topological quantum 
field theories. 

Now we would like to present BF matrix model which has the form
\begin{eqnarray}
S_n^D = \varepsilon^{\mu_1 \mu_2 \cdots \mu_D} 
Tr X_{\mu_1} X_{\mu_2} \cdots X_{\mu_n} B_{\mu_{n+1} \cdots \mu_D},
\label{3.1}
\end{eqnarray}
where a totally antisymmetric tensor matrix $B$ is introduced. 
In this respect let us recall that the original form of 
topological BF theory \cite{BF1, BF2, BF3, BF4} is
\begin{eqnarray}
S_{BF} = \int \varepsilon^{\mu_1 \mu_2 \cdots \mu_D} 
Tr F_{\mu_1 \mu_2} \ B_{\mu_3 \cdots \mu_D},
\label{3.2}
\end{eqnarray}
where the 2-form field strength $F$ is defined as $F = dA + A^2$.
Thus, precisely speaking, the straightforward generalization of the 
topological BF theory (\ref{3.2}) to the matrix model corresponds 
to the case of $n=2$ in (\ref{3.1}). Of course, owing to
the introduction of the matrix $B$ the action (\ref{3.1}) 
makes sense in arbitrary space-time dimension.

The classical equations of motion derived from the BF matrix model 
(\ref{3.1}) read
\begin{eqnarray}
\varepsilon^{\mu_1 \mu_2 \cdots \mu_D} 
X_{\mu_1} X_{\mu_2} \cdots X_{\mu_n} = 0,
\label{3.3}
\end{eqnarray}
\begin{eqnarray}
\sum_{i=1}^n (-1)^{i-1} \varepsilon^{\mu \mu_1 \cdots 
\hat{\mu_i} \cdots \mu_D} X_{\mu_{i+1}} \cdots X_{\mu_n} 
B_{\mu_{n+1} \cdots \mu_D} X_{\mu_1} \cdots X_{\mu_{i-1}}
= 0,
\label{3.4}
\end{eqnarray}
where $\hat{\mu_i}$ denotes that the index $\mu_i$ is excluded.
Note that apart from the number of $X_\mu$ , 
Eq.(\ref{3.3}) accords with (\ref{2.3}) in the Chern-Simons 
matrix theory. Thus the structure of the solutions with respect
to $X_\mu$ is almost the same as that case.
On the other hand, it is Eq.(\ref{3.4}) that appears for the 
first time in the BF matrix model. 
In fact, this equation would have an important 
implication in relating the model at hand to general relativity later.

As for the gauge symmetries, besides the usual U(N) gauge symmetry, 
at first glance the action (\ref{3.1}) looks like it might
be invariant under the following natural geralization of the local 
translation symmetry (\ref{2.11})
\begin{eqnarray}
X_\mu &\rightarrow& X_\mu^\prime =  X_\mu + V_\mu(X) 
\ {\bf 1}, \nn\\
B_{\mu_{n+1} \cdots \mu_D} &\rightarrow& 
B_{\mu_{n+1} \cdots \mu_D}^\prime = B_{\mu_{n+1} \cdots \mu_D}
+ W_{\mu_{n+1} \cdots \mu_D}(X) \ {\bf 1}.
\label{3.5}
\end{eqnarray}
However, it is interesting to notice that only the action 
(\ref{3.1}) with $n$ being even integers has such a local
translation symmetry while the action (\ref{3.1}) with odd $n$
has neither the local nor the global translation symmetry.
Concerning the canonical formalism of the BF matrix model,
although some formulae become more complicated than in
the Chern-Simons matrix model,
the canonical formalism explained
in the previous section applies equally well to this case.
The key point here is to start with the following matrix
model consisting of an almost surface term:
\begin{eqnarray}
I_n^D &=& \int d\tau \varepsilon^{\mu_1 \mu_2 \cdots \mu_D}
Tr \ [ \ ({\it D_\tau} X_{\mu_1}) X_{\mu_2} \cdots 
X_{\mu_n} B_{\mu_{n+1} \cdots \mu_D}         \nn\\
& & {} +  X_{\mu_1}({\it D_\tau} X_{\mu_2}) X_{\mu_3} \cdots 
X_{\mu_n} B_{\mu_{n+1} \cdots \mu_D} + \cdots  \nn\\
& & {} + X_{\mu_1} \cdots X_{\mu_{n-1}}
({\it D_\tau} X_{\mu_n}) B_{\mu_{n+1} \cdots \mu_D} \nn\\
& & {} +  X_{\mu_1} \cdots X_{\mu_n} \bar{{\it D_\tau}} 
B_{\mu_{n+1} \cdots \mu_D} \ ],
\label{3.6}
\end{eqnarray}
where the second covariant derivative 
$\bar{{\it D_\tau}} B_{\mu_{n+1} \cdots \mu_D}
= \partial_\tau B_{\mu_{n+1} \cdots \mu_D} + 
\left[ \bar{A_\tau}, B_{\mu_{n+1} \cdots \mu_D}
\right]$
with another fictitious gauge field $\bar{A_\tau}$
is also introduced.
Following the similar line of arguments to the
case of the Chern-Simons matrix model, we can
also reach the similar results whose details 
are skipped over now. Only the difference
lies in the fact that the trace of both canonical
conjugate momenta to $X_\mu$ and $B_{\mu_{n+1} 
\cdots \mu_D}$ identically vanishes
only for even $n$ which would be related to
the existence of the local translation invariance
for even $n$.

Before closing this section, we turn to the problem of 
relating the present model to general relativity. 
Indeed, there are some methods for it in lower dimensions.
This follows from the fact that the topological 
BF theory includes the content of general relativity
in two, three and four dimensions. 
In contrast, we have no clear understanding of how to
formulate general relativity in terms of the topological
BF theory in the dimensions more than four.    
In this paper, we shall confine our consideration to 
general relativity in four space-time dimensions 
since the treatment 
in both two and three dimensions is also similar to or easier
than in four dimensions.

To this aim, one has to consider the specific case $n=2$,
which exactly corresponds to the matrix model of the
original, topological BF theory. First of all,
let us consider the first possibility of deriving the action
of general relativity in four dimensions by starting with
the BF matrix model in the dimensions more than four.
Then, for a notational convenience, let us decompose the 
index $\mu = 0, 1, \cdots, D-1$ into the four-dimensional 
part $A = 0, 1, 2, 3$ and the remaining part $a=4, \cdots, D-1$.
Moreover, we introduce the definition 
$\tilde{B}^{\mu_1 \mu_2} = \varepsilon^{\mu_1 \mu_2 \cdots \mu_D} 
B_{\mu_3 \cdots \mu_D}$. Consequently, the starting action
is of the form
\begin{eqnarray}
S_{n=2}^D &=& Tr X_{\mu_1} X_{\mu_2} \tilde{B}^{\mu_1 \mu_2} \nn\\
&=& Tr ( X_{A_1} X_{A_2} \tilde{B}^{A_1 A_2} + [ X_A, X_a ]
\tilde{B}^{A a} + X_{a_1} X_{a_2} \tilde{B}^{a_1 a_2} ).
\label{3.7}
\end{eqnarray}
Then, the key idea is to find a special solution satisfying
the equations of motion for (\ref{3.4}) $\tilde{B}$ without 
affecting the space-time coordinates $X_\mu$ in order to yield 
the first-order Palatini action of general relativity
in four dimensions. 
We can easily find the desirable solution given by 
\begin{eqnarray}
\tilde{B}^{A_1 A_2} &=& \varepsilon^{A_1 A_2 A_3 A_4} 
e_{A_3} e_{A_4}, \nn\\
\tilde{B}^{A a} &=& \tilde{b}^{A a} \ {\bf 1}, \
\tilde{B}^{a_1 a_2} = \tilde{b}^{a_1 a_2} \ {\bf 1},
\label{3.8}
\end{eqnarray}
where $e_A$ is the one-form vierbein and 
$\tilde{b}$ is not a matrix but a c-number with the same 
symmetric property as the corresponding matrix. 
If we substitute (\ref{3.8}) into (\ref{3.7}), we can obtain
\begin{eqnarray}
S_{n=2}^D = \varepsilon^{A_1 A_2 A_3 A_4} Tr X_{A_1} X_{A_2} 
e_{A_3} e_{A_4}.
\label{3.9}
\end{eqnarray}
This is exactly the same form of the first-order Palatini action
reduced to a point. 
In this way, we can derive the action of general relativity 
from the BF matrix model in a simple manner. 

Even if the above derivation is itself of interest, some 
people may complain that we have just selected a special solution
by hand among many classical solutions. Here, to make the 
process of selecting the special solution (\ref{3.8}) more 
convincing, we can make use of the strategy adopted in the 
references \cite{CDJ}, which amount to adding an additional 
term to the starting action such that the above solution becomes
the general solution. To elucidate our strategy, let us 
just confine ourselves to four space-time dimensions from
the outset. Then
relevant action equals
\begin{eqnarray}
S_{n=2}^{D=4} 
= Tr ( X_{A_1} X_{A_2} \tilde{B}^{A_1 A_2} 
- \frac{1}{2}  \Psi \tilde{B}^{A_1 A_2} B_{A_1 A_2} ).
\label{3.10}
\end{eqnarray}
The variational equation with respect to $\Psi$ produces 
the equation $\tilde{B}^{A_1 A_2} B_{A_1 A_2} = 0$. According to
the proposition in \cite{Penrose}, the general solution of this
equation is given by $\tilde{B}^{A_1 A_2} = 
\varepsilon^{A_1 A_2 A_3 A_4} e_{A_3} e_{A_4}$. Thus, the 
substitution of this solution into (\ref{3.10}) leads to 
the first-order Palatini action (\ref{3.9}) like before. 
It is quite interesting to examine whether the above-mentioned 
strategy can also be 
applied to the case of the higher space-time dimensions.

\section{ Generalization with fermionic symmetry }

Now we will discuss some generalizations of the BF model
to include fermionic symmetry. Indeed, in the matrix
models \cite{IKKT, M} the fermionic symmetry, in
particular, the supersymmetry, was needed to guarantee
the cluster and BPS properties of instantons. 

One possibility is to add fermions of integer spin
to achieve a BRST-like symmetry. It is known that
the partition function of the BF theory is related
to the Ray-Singer torsion \cite{Schwarz} 
while that of the BF theory with such a BRST-like 
symmetry correspondes to the Casson invariant \cite{Witten2}.
We think that this statement
is valid even in the BF matrix model treated in 
this paper. Let us start by the following BRST-like
fermionic symmetry:
\begin{eqnarray}
\delta X_\mu &=& \eta \psi_\mu, \ \delta \psi_\mu = 0, \nn\\
\delta \chi_{\mu_{n+1} \cdots \mu_D} &=& - \eta B_{\mu_{n+1} 
\cdots \mu_D}, \ \delta B_{\mu_{n+1} \cdots \mu_D} = 0.
\label{4.1}
\end{eqnarray}
We can check explicitly the following action to be invariant
under the fermionic symmetry (\ref{4.1}):
\begin{eqnarray}
S_n^D &=& \varepsilon^{\mu_1 \mu_2 \cdots \mu_D} 
Tr ( X_{\mu_1} X_{\mu_2} \cdots X_{\mu_n} B_{\mu_{n+1} 
\cdots \mu_D} \nn\\
& & {} - \sum_{i=1}^n X_{\mu_1} X_{\mu_2} \cdots X_{\mu_{i-1}}
\psi_{\mu_i} X_{\mu_{i+1}} \cdots X_{\mu_n} \chi_{\mu_{n+1}
\cdots \mu_D} )
\label{4.2}
\end{eqnarray}
{}For even integers $n$, this action is still invariant under 
the enlarged local translation which constitutes of Eq.(\ref{3.5}) 
and
\begin{eqnarray}
\psi_\mu &\rightarrow& \psi_\mu^\prime =  \psi_\mu + v_\mu(X)
\ {\bf 1}, \nn\\
\chi_{\mu_{n+1} \cdots \mu_D} &\rightarrow& 
\chi_{\mu_{n+1} \cdots \mu_D}^\prime = \chi_{\mu_{n+1} \cdots \mu_D}
+ w_{\mu_{n+1} \cdots \mu_D}(X) \ {\bf 1}.
\label{4.3}
\end{eqnarray}

A more interesting possibility of incorporaing fermions of half
integer spin would be to twist the action (\ref{4.2}) like the
topological quantum field theory \cite{Witten3}. Here note that 
even if the bosonic action
(\ref{3.1}) is nontrivial its BRST-like generalization (\ref{4.2})
is BRST-exact form so that we can use the twisting technique
developed in the reference \cite{Witten3}. This problem will be
reported in a separate publication in future.

\section{ Discussions }
In this paper, we have proposed two candidates for the background
independent formulation of the matrix model. One is based on
the Chern-Simons theory \cite{Witten1} in odd dimensions
\cite{Smolin}, and the other is on the BF 
theory \cite{BF1, BF2, BF3, BF4} in any space-time 
dimension. Both of the models share some common features, for 
instance, the existence of similar gauge symmetries and classical
solutions. However, it seems that the latter matrix model is
currently more interesting than the former one in that the 
BF matrix model not only can be formulated in an arbitrary 
dimension, in particular, 
in four dimensions, but also has a close connection with gravity.
Note that these advantageous features of the BF theory are already 
seen at the level of topological field theory.      
   
As mentioned in the introduction, the main purpose of constructing
the background independent matrix models is to understand the
non-perturbative aspects of string theory without reference to
the specific background metric. To this end, we have to
clarify the relation in detail between the present matrix models
and the ones in \cite{IKKT, M} in future. At any rate, it
seems to be essential to twist the model at hand for the purpose
of getting $N=2$ supersymmetric matrix model.  

\vs 1
\begin{flushleft}
{\bf Acknowledgement}
\end{flushleft}
The author is grateful to A.Sugamoto for valuable discussions
and continuous encouragement. 
This work was supported in part by Grant-Aid 
for Scientific Research from Ministry of Education, Science and
Culture No.09740212.

\vs 1

\end{document}